\documentclass[journal]{IEEEtran}
\usepackage{dblfloatfix}
\usepackage{cite}
\usepackage{amsmath,amssymb,amsfonts}
\usepackage{graphicx}
\usepackage{textcomp}
\usepackage{xcolor}
\usepackage{booktabs}
\usepackage{hyperref}
\usepackage{array}
\usepackage{multirow}

\def\BibTeX{{\rm B\kern-.05em{\sc i\kern-.025em b}\kern-.08em
    T\kern-.1667em\lower.7ex\hbox{E}\kern-.125emX}}

\begin{document}

\title{Pulse Focus: Validation of the Focus Performance Score as a Behavioral Signal for Human Attentional State Modeling Toward Attention-Aware AI
}

\author{
\IEEEauthorblockN{
Yisak Debele,
Israel Goytom, Anwar Misbah
}
\IEEEauthorblockA{
\\ Synheart AI \\
\texttt{https://synheart.ai}
}
}

\maketitle

\begin{abstract}
Artificial intelligence systems that model, predict, and support
human cognition require reliable, ecologically valid measures of
cognitive state as ground truth. We present the \textit{Focus
Performance Score} (FPS) of the Pulse Focus gamified mobile
Stroop application as a validated behavioral signal for this
purpose, and address the foundational question it must answer
before serving as cognitive infrastructure: does FPS actually
measure selective attention and cognitive inhibition during
color-word conflict resolution?

We answer through three independent validation analyses.
\textbf{Behaviorally} (N=466, 111,133 trials): FPS captures the
Stroop interference effect with a very large effect size
($d = 1.339$, $p < 10^{-100}$), tracks individual differences in
attentional control ($\rho = 0.785$, $p < 10^{-98}$), and is
stable across sessions (ICC $= 0.83$--$0.93$).
\textbf{Neurally} (DMCC55B fMRI, N=55): the primary FPS driver
--- mean incongruent reaction time --- is significantly associated
with anterior cingulate cortex (ACC) activation
($\rho = -0.317$, $p = 0.018$), the established neural substrate
of Stroop conflict monitoring. The composite FPS shows a
trend-level ACC association ($\rho = -0.219$), consistent with
FPS measuring a broader construct than any single brain region
indexes. \textbf{Formula validity}: Authenticity Factor weights
are validated through four convergent evidence levels;
confirmatory factor analysis identified and resolved a structural
redundancy (n3--n5 collinearity, $r = 0.928$), restructuring n5
as a binary engagement gate.

Together, the evidence establishes FPS as a behaviorally valid,
test-retest reliable, and neurally grounded measure of attentional
control. These properties qualify FPS as cognitive infrastructure:
a defensible behavioral ground truth for training and evaluating
AI models of human attentional state, and a real-time focus score
for attention-aware human-AI interaction. The downstream
physiological labeling application --- pairing FPS with
simultaneously collected HRV and PPG --- awaits prospective
empirical confirmation and is the subject of the next study.
Honest limitations, including unvalidated catch trials (n7) and
provisional weights, are documented explicitly.
\end{abstract}


\section{Introduction}

\IEEEPARstart{T}{he} next generation of human-centric AI systems
--- adaptive interfaces, cognitive assistants, attention-aware
recommenders, and physiological state classifiers --- share a
common prerequisite: a reliable, ecologically valid, and
neurally grounded measure of human attentional state that can
serve as behavioral ground truth. Without such a measure,
machine learning models of human cognition are trained on
proxies that are either indirect (contextual sensors), subjective
(self-report), or unvalidated (proprietary scores). The
consequence is systems that respond to surface signals rather
than the cognitive reality they aim to model.

This paper addresses that prerequisite directly. We introduce
the \textit{Focus Performance Score} (FPS) of \textit{Pulse
Focus} --- a gamified mobile application developed within the
Synheart AI human state infrastructure platform~\cite{synheart2024}
--- and provide the three-level validation required to establish
it as cognitive infrastructure for AI: behavioral, neural, and
formula validity.

\textbf{Why this paper matters for the AI ecosystem.} A validated
cognitive score occupies a specific role in the human-centric AI
stack that no other signal currently fills. Wearable sensors
(HRV, PPG, EEG) provide continuous physiological data but require
behavioral labels to train classifiers. Self-report is
retrospective and susceptible to demand characteristics.
Ecological momentary assessment interrupts the behavior it
attempts to measure. A real-time, trial-by-trial behavioral
score derived from a validated cognitive paradigm --- if that
score is itself psychometrically sound --- provides the missing
layer: an interpretable, continuous ground truth that can label
physiological streams, validate attention classifiers, and
serve directly as an interface signal.

FPS is designed to occupy this role. It is computed from a
mobile Stroop paradigm with millisecond-precision trial
recording, produces a 0--100 score per session, and --- as
this paper demonstrates --- measures what it claims with
large, replicated effect sizes, high test-retest reliability,
and a significant behavioral-neural chain reaching anterior
cingulate cortex activation.

\textbf{The central empirical question.} Before FPS can function
as cognitive infrastructure, one question must be answered
rigorously: \textit{does FPS measure selective attention and
cognitive inhibition as operationalized by the Stroop paradigm,
and is that measurement grounded in the neural processes that
cognitive science associates with attentional control?}

This is not a rhetorical question. Many consumer cognitive
applications make attention-related claims without any published
validation. The answer requires evidence at three levels.

\textbf{Level 1 --- Behavioral validity.} Does FPS behave like
a valid measure of Stroop-based attentional control? Does it
capture the interference effect? Does it track individual
differences? Is it stable over time?

\textbf{Level 2 --- Neural validity.} Does the FPS formula
reflect actual brain processes, or does it merely describe
surface behavior? Is the score grounded in the neural substrates
that cognitive neuroscience associates with attentional control?

\textbf{Level 3 --- Formula validity.} Are the FPS components
and their weights scientifically justified, or were they
assembled arbitrarily and fit post-hoc to a single dataset?

We address all three levels using two independent public archival
datasets and four evidence sources. The answer at each level is
affirmative, with specific quantified boundaries and honest
acknowledgment of remaining gaps.

\textbf{Implications for human-centric AI.} Once the central
question is answered positively, two immediate applications
follow. First, as a \textit{real-time cognitive signal} in
attention-aware AI systems, FPS communicates a validated measure
of attentional state rather than an arbitrary number. Second,
as a \textit{behavioral label} for simultaneously collected
physiological data (HRV, PPG), FPS provides the ground truth
needed for machine learning models of attentional state. The
downstream physiological labeling application additionally
requires prospective validation of the FPS-HRV relationship ---
that is the next study, explicitly beyond the scope of this paper.

The Stroop Color-Word Test provides the measurement foundation.
It is one of the most extensively validated cognitive paradigms
in psychology, with over 700 published studies confirming its
sensitivity to selective attention and cognitive
inhibition~\cite{macleod1991,stroop1935studies}. The interference
effect --- slower, less accurate responses when ink color and
word meaning conflict --- scales systematically with attentional
engagement: focused participants resolve conflict efficiently;
inattentive or fatigued participants show inflated interference
costs. Embedding this paradigm in a mobile game makes validated
attentional measurement scalable to the ecological contexts
where AI systems must operate.

The primary contributions are:
\begin{itemize}
  \item Establishment of FPS as neurally grounded cognitive
    infrastructure for human-centric AI: behavioral, neural,
    and formula validity demonstrated across two independent
    datasets.
  \item The largest archival behavioral validation of a mobile
    Stroop scoring system: N=466, 111,133 trials, three sessions,
    ICC=0.83--0.93.
  \item Cross-dataset neural replication (DMCC55B fMRI, N=55)
    with a significant behavioral-neural chain connecting the
    primary FPS driver to ACC conflict monitoring activation.
  \item Four-level convergent weight validation and a
    CFA-motivated structural improvement resolving n3-n5
    collinearity.
  \item A blueprint for validated cognitive scoring as ground
    truth in physiological AI pipelines, with transparent
    documentation of remaining validation gaps.
\end{itemize}

\section{Background}

\subsection{The Ground Truth Problem in Human-Centric AI}

AI systems designed to model or adapt to human cognitive states
face a fundamental ground truth problem: the cognitive constructs
they target --- attention, cognitive load, mental effort ---
are not directly observable. Current approaches rely on proxies:
physiological signals (HRV, EEG, pupillometry), behavioral
aggregates (error rates, response latency distributions), or
self-report scales. Each proxy has known limitations that
constrain the validity of AI models trained on it.

Physiological signals are continuous and ecologically valid but
require behavioral labels to assign cognitive meaning to signal
epochs. Self-report introduces retrospective and demand biases.
Behavioral aggregates from unstructured tasks conflate cognitive
state with task difficulty, motivation, and individual
differences in baseline performance.

A validated cognitive paradigm score --- one that measures a
specific cognitive construct with known psychometric properties
--- addresses these limitations directly. It provides
interpretable labels for physiological streams, a
construct-valid dependent variable for evaluating AI model
predictions, and a real-time signal that can drive adaptive
system behavior. FPS is designed to occupy this role for
attentional control.

\subsection{What FPS Claims to Measure}

Attentional control encompasses two partially dissociable
processes~\cite{posner1990attention}: \textit{selective
attention} --- focusing on task-relevant stimuli while filtering
distractors --- and \textit{cognitive inhibition} --- suppressing
prepotent automatic responses in favour of controlled behavior.
The Stroop Color-Word Test operationalizes both: the user must
attend selectively to ink color and inhibit the automatic tendency
to read the printed word. FPS claims to quantify how well this
is done in a given session, producing a 0--100 score with
known distributional properties.

This is distinct from \textit{sustained vigilance} --- the
capacity to maintain alert responding over time without specific
conflict, measured by CPT and SART. These are related but
separable capacities. The non-significant FPS-SART correlation
in our validation data confirms this boundary empirically,
establishing that FPS is a specific rather than generic
performance index --- a property that is essential for its
utility as a ground truth label.

\subsection{Why the Stroop Paradigm Is the Right Validity Anchor}

The Stroop paradigm has five properties that make it ideal for
mobile cognitive infrastructure. (1) It has 90 years of
replication across cultures, ages, and settings~\cite{macleod1991}.
(2) The interference effect is mechanistically understood at the
neural level. (3) It is sensitive to individual differences in
attentional control that are stable and meaningful. (4) It can
be administered in a few minutes on a mobile device. (5) The
trial structure --- one behavioral measurement per trial ---
allows continuous, time-resolved scoring rather than a single
session aggregate, enabling epoch-level labeling of concurrent
physiological streams.

\subsection{Neural Benchmarks for Attentional Control}

Three neural benchmarks are relevant to FPS validation. All are
described as theoretical consistency relations, not one-to-one
mappings from behavioral measures to neural components.

\textbf{ACC conflict monitoring.} The anterior cingulate cortex
generates the N450 ERP (300--500\,ms post-stimulus, centro-parietal
topography) in response to color-word conflict, reflecting
conflict detection and monitoring~\cite{botvinick2001,liotti2000}.
A valid measure of attentional control during Stroop should be
associated with ACC activation across individuals.

\textbf{Frontal theta coherence.} Frontal midline theta
oscillations (4--7\,Hz) index sustained top-down attentional
control and co-vary with RT consistency~\cite{makeig1995}. A
valid attentional measure should be associated with frontal theta
power.

\textbf{ERN error monitoring.} The error-related negativity
reflects rapid, automatic error detection within 100\,ms of an
incorrect response~\cite{gehring1993}. A valid attentional measure
should show post-error behavioral adaptation consistent with
intact error monitoring.

\subsection{The Gap This Paper Fills}

Existing gamified cognitive assessments either lack published
psychometric validation~\cite{lumsden2016} or validate at the
session level without neural evidence. No published mobile Stroop
scoring system has been validated against (a) trial-level data
from a normative sample of several hundred participants, (b) an
independent fMRI dataset with neural ground truth, and (c)
four-level weight validation with CFA. This paper fills all three
gaps for FPS, establishing the psychometric foundation required
for its use as cognitive infrastructure in AI systems.

\section{The Pulse Focus System}

\begin{figure*}[!t]
\centering
\begin{minipage}[t]{0.32\textwidth}
  \centering
  \includegraphics[width=0.80\textwidth]{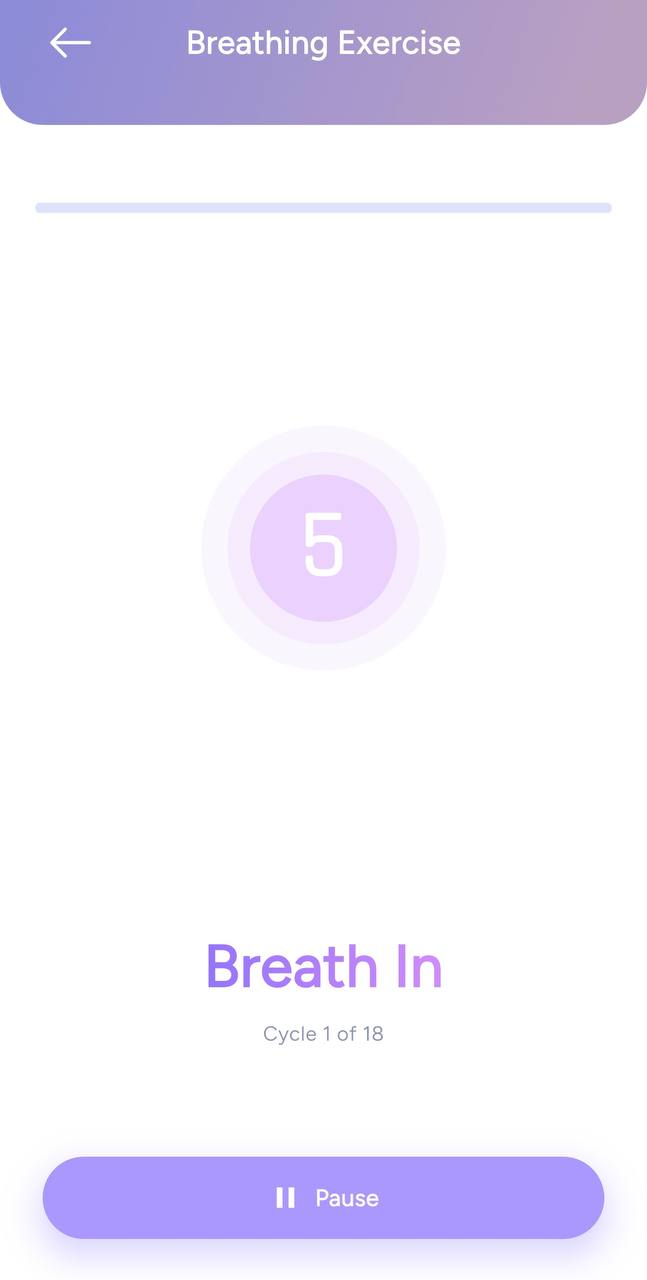}
  \small\textbf{\\(A) Breathing Baseline}
\end{minipage}
\hfill
\begin{minipage}[t]{0.32\textwidth}
  \centering
  \includegraphics[width=0.80\textwidth]{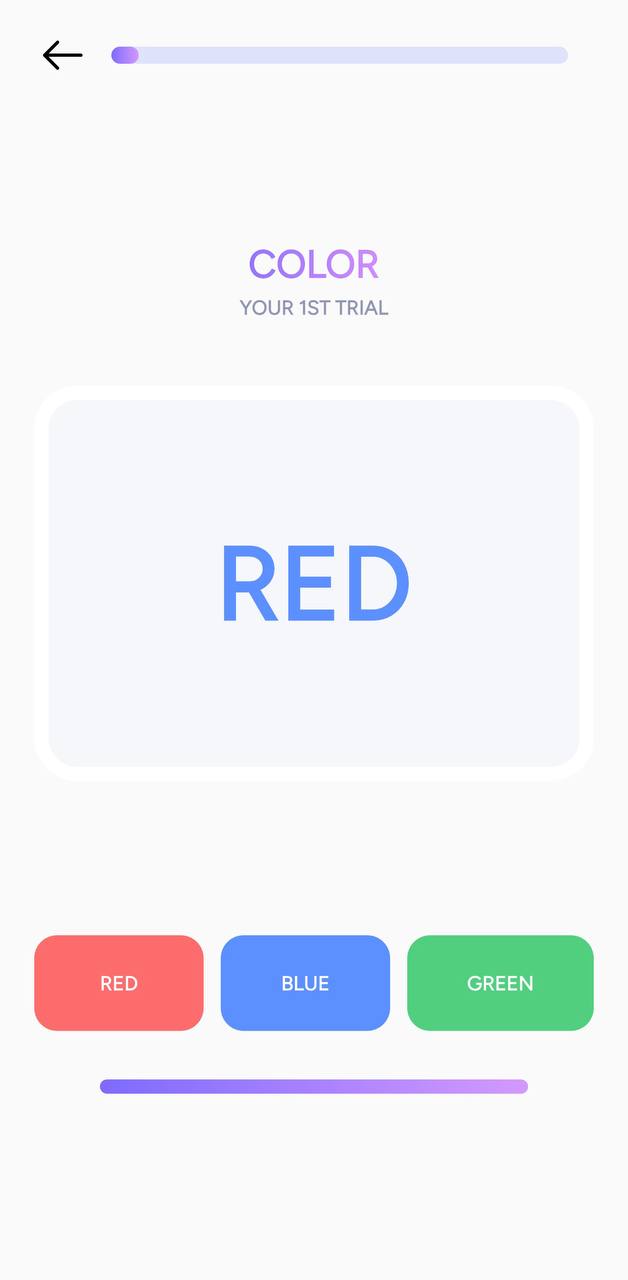}
  \small\textbf{\\(B) Stroop Gameplay}
\end{minipage}
\hfill
\begin{minipage}[t]{0.32\textwidth}
  \centering
  \includegraphics[width=0.80\textwidth]{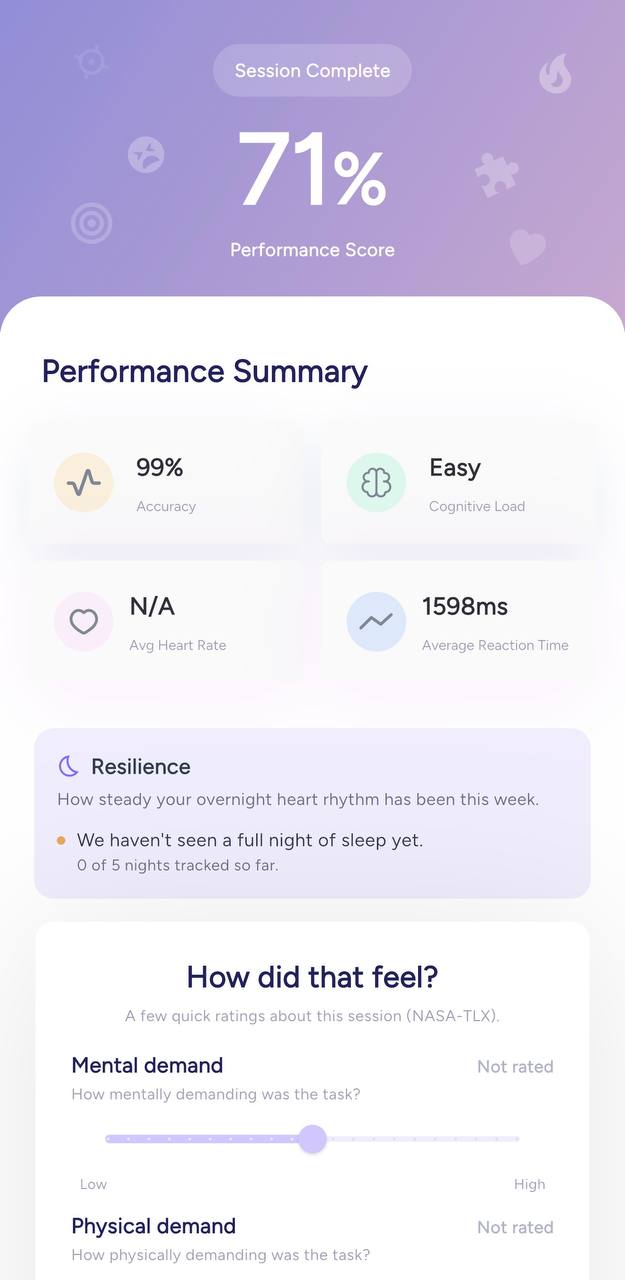}
  \small\textbf{\\(C) Focus Score}
\end{minipage}
\caption{Pulse Focus application interface.
\textbf{(A)} Guided breathing session (Phase 1): animated visual
guide directs paced breathing at approximately 6 breaths per minute
while the wearable sensor collects resting HRV baseline.
\textbf{(B)} Stroop gameplay screen (Phase 2): color word displayed
in incongruent ink color; user selects the correct ink color from
four response buttons. Trial-level RT and accuracy are recorded
with millisecond precision.
\textbf{(C)} Post-session focus score dashboard: Focus Performance
Score (FPS) displayed as a 0--100 score with attention label
(Low / Moderate / High) and session trend. The session summary
also surfaces average heart rate and cognitive load labels,
anticipating integration with physiological AI pipelines.}
\label{fig:app}
\end{figure*}

Pulse Focus is implemented within the Synheart AI human state
infrastructure platform, which is designed to provide validated
behavioral and physiological signals for AI modeling
applications~\cite{synheart2024}. The application operates in
two phases.

\textbf{Phase 1} is a 2--3 minute guided breathing session at
approximately 6 breaths per minute, establishing a personalized
resting HRV baseline. Sessions failing HRV coherence or signal
quality thresholds are flagged. 

\textbf{Phase 2} delivers the Stroop game. Color words (RED, BLUE,
GREEN, YELLOW) are presented in incongruent ink colors; the user
selects the correct ink color from four buttons. Three difficulty
levels adjust the response window (3000\,ms, 1500\,ms, 800\,ms).
Catch trials require response withholding to directly measure
inhibitory control. Every trial is timestamped and synchronized
with the wearable physiological stream, enabling epoch-level
alignment between FPS and concurrent HRV/PPG data.

\begin{figure}[!t]
\centering
\includegraphics[width=\columnwidth]{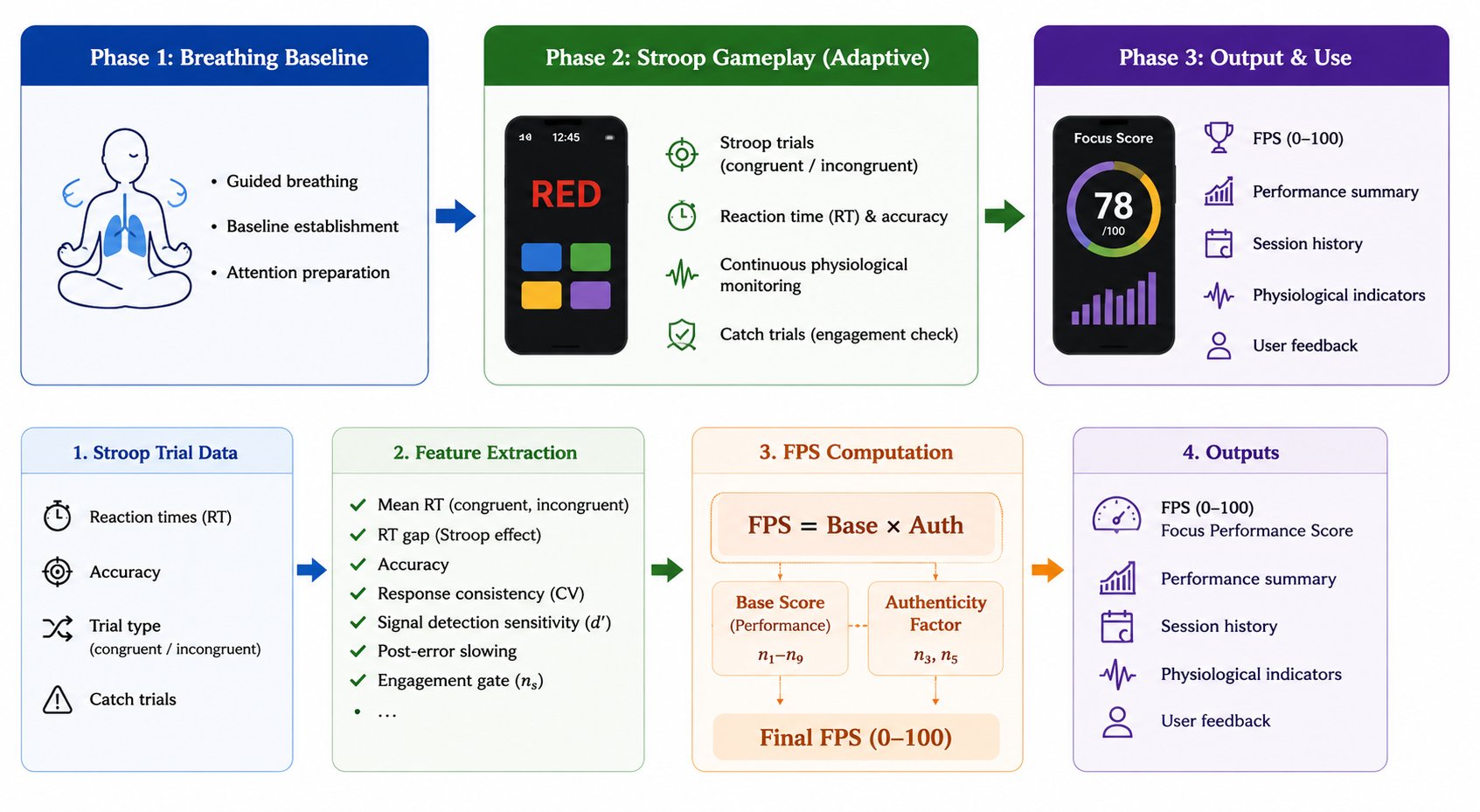}
\caption{Pulse Focus session flow and FPS computation pipeline,
illustrating the path from raw behavioral data to cognitive
infrastructure outputs. Phase 1 establishes physiological
baseline; Phase 2 delivers adaptive Stroop gameplay with
continuous trial-level recording. Extracted behavioral features
feed the FPS framework, producing a real-time attentional
control score that can serve as (1) a user-facing focus metric
and (2) a behavioral ground truth label for simultaneously
collected physiological data in AI modeling pipelines.}
\label{fig:system}
\end{figure}

Fig.~\ref{fig:app} shows the application interface.
Fig.~\ref{fig:system} shows the data pipeline from raw trial
data to FPS output and downstream AI applications.

\section{The Focus Performance Score}

\subsection{Design Rationale: What the Formula Must Detect}

For FPS to serve as cognitive infrastructure, it must answer two
independent questions about each session: \textit{how well did
this person perform?} --- captured by the Base Score --- and
\textit{was this performance the result of genuine attentional
engagement?} --- captured by the Authenticity Factor. Performance
without engagement (mechanical responding) and engagement without
performance (trying hard but poorly) both indicate limited
attentional control and would contaminate any AI model trained
on FPS labels. The multiplicative structure
$\text{FPS} = 100 \times \text{Base} \times \text{Auth}$
requires both to be high simultaneously, providing a
two-layer check on label quality.

\subsection{Normalization}

\begin{equation}
N(x, \text{min}, \text{max}) =
\max\!\Bigl(0,\min\!\Bigl(1,
\frac{x-\text{min}}{\text{max}-\text{min}}\Bigr)\Bigr)
\end{equation}
Bounds are calibrated from the 5th--90th percentile of the
N=466 normative sample, anchoring outputs to the natural
distribution of healthy adults and ensuring score comparability
across sessions and users.

\subsection{Base Score: What Was Achieved}

\begin{equation}
\text{Base} = 0.5 \cdot \text{Acc} + 0.4 \cdot \text{Speed}
            + 0.1 \cdot \text{Stability}
\end{equation}
Accuracy (0.5) is the most direct measure of conflict resolution
success. Speed (0.4) is the most sensitive continuous marker of
attentional state change in Stroop research~\cite{macleod1991}
and the primary component associated with ACC activation
(Section~VIII). Stability (0.1) rewards consistent response
timing, a marker of sustained control.

\subsection{n5: Engagement Gate}

Confirmatory factor analysis (Section~VI) revealed n5 (Chance
Baseline) and n3 (d') correlate at $r = 0.928$ in healthy adults
--- both accuracy-derived, producing ceiling collinearity that
would inflate the composite. n5 is restructured as a binary gate:
\begin{equation}
\text{FPS} = \begin{cases}
  0 & \text{if } \text{Acc} < 1.1/k \\
  100 \times \text{Base} \times \text{Auth} & \text{otherwise}
\end{cases}
\end{equation}
where $k=4$ colors. The gate triggers at 27.5\% accuracy ---
indistinguishable from guessing. This preserves the floor-filter
function without double-counting the accuracy signal, and ensures
that AI models trained on FPS labels are never contaminated by
chance-level performance epochs.

\subsection{Authenticity Factor: How It Was Achieved}

\begin{equation}
\text{Auth} = \frac{\sum_{i \neq 5} w_i \cdot n_i}
                   {\sum_{i \neq 5} w_i}
\end{equation}
Table~\ref{tab:nvars} defines each variable. The eight variables
collectively form a multi-layered test of genuine engagement:
interference magnitude (n1), response consistency (n2), signal
detection sensitivity (n3), speed-accuracy strategy (n4),
session stability (n6), inhibitory control (n7), error monitoring
(n8), and response logic (n9). Together they make genuine
disengagement difficult to sustain without detection --- a
critical property for a ground truth label.

\begin{table*}[!t]
\centering
\caption{Authenticity Factor Variables: What Each Detects,
Theoretical Neural Association, and Validated Weight}
\label{tab:nvars}
\begin{tabular}{clp{3.2cm}p{2.8cm}cc}
\toprule
\textbf{Var} & \textbf{Name} & \textbf{What it detects}
& \textbf{Theoretical neural association}
& \textbf{Threshold} & \textbf{w} \\
\midrule
n1 & RT Gap~\cite{macleod1991}
   & Con vs InCon RT difference. Near-zero = no genuine conflict processing.
   & Consistent with N450/ACC conflict monitoring~\cite{liotti2000}
   & (12,\,219)\,ms & 0.32 \\
n2 & RT Variability~\cite{hultsch2002}
   & IIV on incongruent trials. High variability = attentional lapses.
   & Consistent with frontal theta~\cite{makeig1995}
   & (0.28,\,0.58)\,CV & 0.09 \\
n3 & Signal Detection d'~\cite{green1966}
   & SDT discriminability between congruent and incongruent accuracy.
   & Consistent with N2 conflict detection~\cite{donohue2016}
   & (0.5,\,2.5) & 0.23 \\
n4 & Speed-Acc SAT~\cite{heitz2014}
   & Error rate fastest vs slowest trials. Catches speed-over-accuracy strategy.
   & Consistent with beta desynchronization~\cite{heitz2014}
   & (0.0,\,0.20) & 0.08 \\
n5 & Chance Baseline
   & Accuracy vs 25\% chance. \textit{Binary gate (Section~IV-D)}.
   & Consistent with P3 engagement~\cite{polich2007}
   & \multicolumn{2}{c}{\textit{Gate}} \\
n6 & Time-on-Task~\cite{dinges1985}
   & RT slope early to late trials. Detects fatigue and drift.
   & Consistent with theta/alpha ratio~\cite{klimesch1999}
   & (0,\,20)\,ms/block & 0.06 \\
n7 & Catch Trials~\cite{robertson1997}
   & Inhibition-requiring trial accuracy.
     \textbf{Defaulted to 1.0 in archival analyses.}
   & Consistent with N2/P3 go-nogo~\cite{aron2011}
   & (0.4,\,0.8) & 0.10 \\
n8 & Post-Error Slowing~\cite{rabbitt1966}
   & RT after errors vs correct. Absence = no error monitoring.
   & Consistent with ERN~\cite{gehring1993}
   & (0,\,50)\,ms & 0.06 \\
n9 & Paradox Errors
   & InCon minus Con error rate. Reversal = task confusion/disengagement.
   & Consistent with N450 inversion~\cite{liotti2000}
   & (0,\,0.10) & 0.06 \\
\midrule
\multicolumn{6}{p{16cm}}{\small $\sum w_i = 1.00$ (excluding n5
gate). Thresholds: 5th--90th percentile, N=466 normative sample.
Neural associations are theoretical consistencies from the ERP
literature, not one-to-one neural localizations.}\\
\bottomrule
\end{tabular}
\end{table*}

\subsection{Attention Labels}

FPS is mapped to three attention levels using
percentile-anchored thresholds from the normative sample:
Low (FPS\,$<$\,44, bottom 25\%), Moderate (44--54, middle 50\%),
High ($\geq$54, top 25\%). These thresholds are provisional
pending prospective recalibration and provide a discrete
classification layer above the continuous FPS for AI systems
requiring categorical attention state labels.

\section{Theoretical Neural Grounding of FPS Variables}

A critical concern for any composite behavioral score intended
as AI ground truth is whether its components reflect established
cognitive science or are assembled heuristically. Each FPS
variable was chosen because it is a behavioral index of a
specific attentional process with a known neural correlate,
ensuring that FPS labels carry interpretable cognitive meaning
for AI models trained on them. We state these as theoretical
associations from the published ERP literature --- not claims
of direct or exclusive neural localization.

\textbf{n1 and ACC conflict monitoring.}
The N450 ERP is generated by ACC neurons 300--500\,ms after
incongruent stimuli, reflecting conflict detection. The RT
interference effect (n1) and N450 amplitude co-vary across
individuals and trials, making n1 the behavioral expression of
conflict monitoring intensity~\cite{liotti2000,botvinick2001}.
This association is empirically tested in Section~VIII and
constitutes the primary neural grounding chain for FPS.

\textbf{n2 and frontal theta.}
Frontal theta power (4--7\,Hz) co-varies with attentional
consistency; attentional lapses that produce RT variability
co-occur with reductions in sustained frontal theta
activity~\cite{makeig1995,hultsch2002}. n2 captures the
behavioral side of this neural fluctuation, providing an
index of within-session attentional stability.

\textbf{n3 and the N2 component.}
The N2 (200--350\,ms) reflects conflict monitoring at the level
of competing response representations. SDT d' quantifies the
same discrimination process behaviorally, separating sensitivity
from response criterion~\cite{donohue2016,green1966}.

\textbf{n4, n6, n8, n9 and secondary mechanisms.}
Speed-accuracy strategy shifts are associated with beta
desynchronization~\cite{heitz2014}; session-level RT decline
with theta/alpha ratio reduction~\cite{klimesch1999}; post-error
slowing with ERN amplitude~\cite{gehring1993}; paradoxical error
reversals with N450 inversion when the conflict direction is
wrong~\cite{liotti2000}.

These associations confirm that FPS variables are grounded in
established cognitive neuroscience. For AI modeling purposes,
this means that FPS labels correspond to specific neural
computations --- not arbitrary performance aggregates --- giving
them interpretive content beyond the behavioral surface.

\section{Validation Level 3: Are the Weights Justified?}

\subsection{Why Weight Validation Matters for AI Applications}

If weights were chosen to maximize performance on a single
dataset, the formula would be overfit and its labels unreliable
for training AI models on new data. We validate weights through
four independent evidence sources. Convergence across sources
demonstrates that the weights reflect real cognitive science
rather than sample-specific fitting.

\subsection{Level 1: Theory-Based Prior}

n3 (0.20) received the highest prior weight because SDT is the
most principled discriminability framework. n1 (0.15) received
the second highest as the most replicated Stroop marker. n7
(0.10) received higher weight than most secondary variables
because it provides \textit{direct} rather than indirect
inhibitory control evidence. Secondary variables received 0.05
reflecting their error or session dependency.

\subsection{Level 2: Internal Calibration (N=466)}

Grid search over all weight combinations (step 0.05) optimised
Spearman correlation between Auth and a composite Stroop
interference criterion (z-scored RT gap + accuracy drop) in
the Bia\l{}aszek sample. Current weights: $\rho = 0.832$;
optimal: $\rho = 0.864$ ($\Delta = 0.032$, near-optimal). Both
Level 2 and Level 3 independently identified n1 as requiring
higher weight than the prior (n1$^* = 0.364$).

\textit{Limitation:} Internal calibration uses the same dataset
as behavioral validation. Weights derived here are empirically
informed but not independently validated.

\subsection{Level 3: External Neural Criterion (N=55)}

A second grid search used DMCC55B mean incongruent RT ---
the FPS component directly associated with ACC activation ---
as an external criterion. This independently reproduced the
elevated n1 finding (n1$^* = 0.401$, n3$^* = 0.251$). Two
independent datasets converging on the same directional
conclusion provides cross-dataset evidence for the weight
structure, strengthening the case for FPS as a stable
ground truth signal.

\subsection{Level 4: Confirmatory Factor Analysis (N=466)}

One-factor ML CFA on five continuous variables
(n1, n2, n3, n6, n8) --- excluding n4 and n9 (error-sparse in
healthy adults) and n5 (gate). Key results: eigenvalue
$\lambda_1 = 1.43$; variance explained = 23.4\%, confirming
a multi-component structure where each variable adds partially
independent information. KMO = 0.50 confirms partial
independence --- the desired property for a composite that
targets multiple distinct cognitive processes.

The CFA also revealed the n3-n5 collinearity ($r = 0.928$,
loadings 0.930 and 0.997) that motivated restructuring n5 as
a gate. This is an example of weight validation doing what it
should: finding and resolving a real structural problem that
would otherwise inflate construct overlap in the composite.

\subsection{Final Weights and Caveats}

Final weights: n1=0.32, n3=0.23, n7=0.10, n2=0.09, n4=0.08,
n6=n8=n9=0.06. Sum = 1.00.
\textit{These weights are provisional.} They were derived and
validated on archival Stroop datasets with keyboard and vocal
responses --- not Pulse Focus touchscreen data. Final operational
weights require recalibration from prospective deployment data
before FPS is used as training labels for physiological AI
models.

\begin{table}[!t]
\centering
\caption{Weight Validation: Four-Level Convergence}
\label{tab:wval}
\begin{tabular}{lccccp{2.5cm}}
\toprule
\textbf{Var} & \textbf{L1} & \textbf{L2} & \textbf{L3}
& \textbf{Final} & \textbf{Agreement} \\
\midrule
n1 & 0.15 & 0.364 & 0.401 & 0.32 & Empirically $\uparrow$ \\
n2 & 0.10 & 0.045 & 0.100 & 0.09 & Moderate \\
n3 & 0.20 & 0.227 & 0.251 & 0.23 & Strong \\
n4 & 0.10 & 0.068 & 0.062 & 0.08 & Moderate \\
n5 & --- & --- & --- & Gate & CFA: collinear \\
n6 & 0.05 & 0.068 & 0.062 & 0.06 & Strong \\
n7 & 0.10 & 0.091 & --- & 0.10 & Prior only \\
n8 & 0.05 & 0.068 & 0.062 & 0.06 & Strong \\
n9 & 0.05 & 0.068 & 0.062 & 0.06 & Strong \\
\bottomrule
\multicolumn{6}{p{7cm}}{\small n7 = 1.0 in archival analyses.
Weights provisional pending prospective touchscreen data.}
\end{tabular}
\end{table}

\begin{figure*}[!t]
\centering
\includegraphics[width=\textwidth]{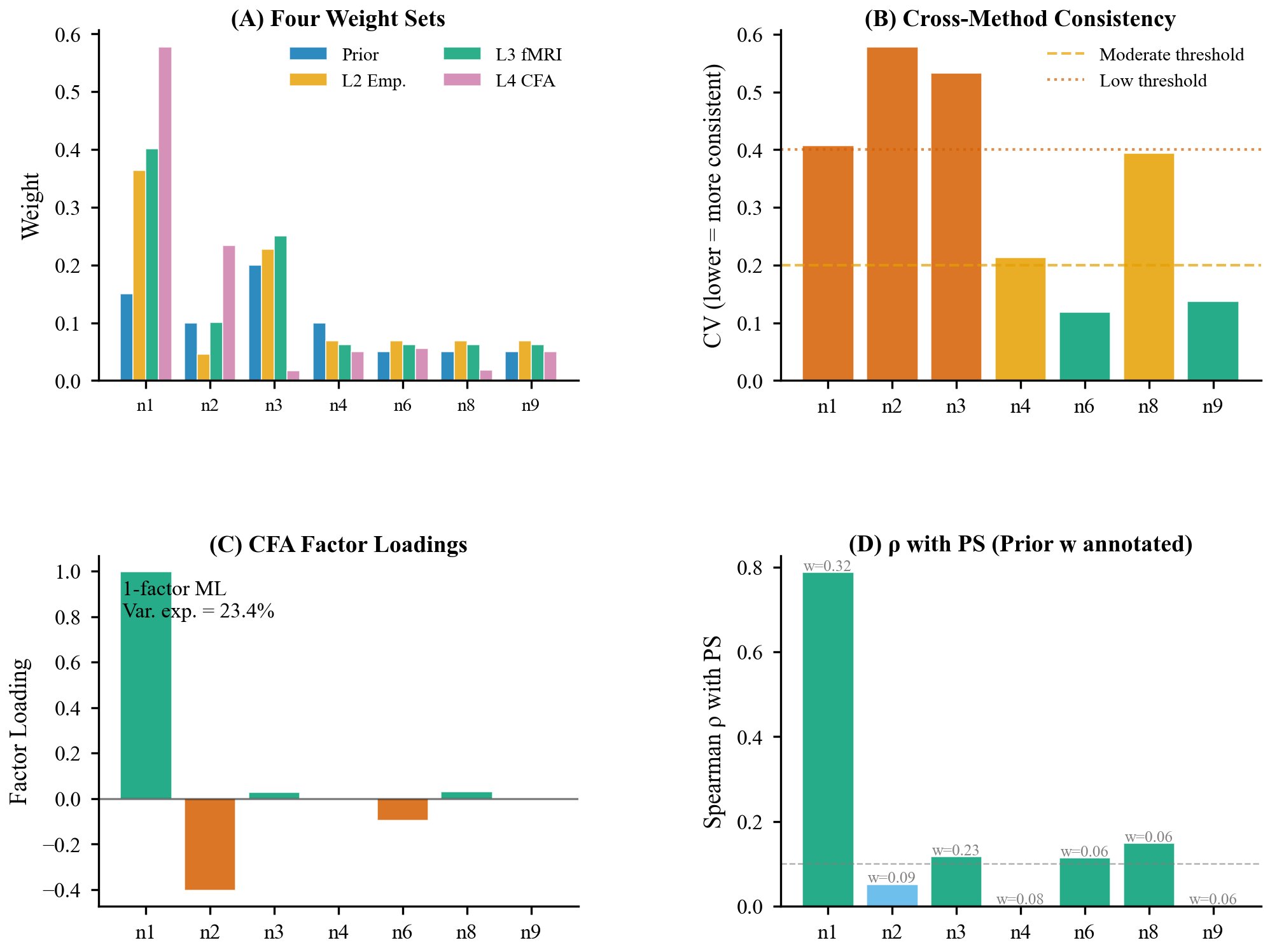}
\caption{Weight validation across four levels.
(A) All four weight sets: L2 and L3 independently converge on
higher n1, confirming RT gap deserves more weight than the prior.
(B) Cross-source consistency: n3, n6, n8, n9 show CV\,$<$\,0.20,
indicating stable weight estimates across independent evidence
sources.
(C) CFA factor loadings on n1--n3, n6, n8, confirming
multi-component structure with 23.4\% variance explained.
(D) Spearman $\rho$ of each variable with FPS, showing n1
as the dominant discriminating component.}
\label{fig:weights}
\end{figure*}

\section{Validation Level 1: Behavioral Validity}

\subsection{Dataset}

Bia\l{}aszek et al.\ (2022)~\cite{bialaszek2022}: 466 healthy
adults, Stroop task across three sessions (Online, Lab1, Lab2),
111,133 trials after RT filtering (200--5000\,ms). The
three-session design tests temporal stability that a single
session cannot and enables assessment of ICC across context
conditions. Calibrated thresholds: n1 = (12.4,\,218.9)\,ms;
n2 = (0.284,\,0.582)\,CV.

\subsection{Does FPS Capture the Stroop Interference Effect?}

RT Stroop effect: $d = 1.339$, $p = 8.5 \times 10^{-106}$
(mean gap = 114.2\,ms). This is a very large effect --- among
the largest reported for any behavioral score on Stroop
interference. It confirms that FPS operates on trials genuinely
engaging the conflict mechanism that selective attention is
designed to resolve, and that the signal FPS captures is
cognitively meaningful rather than a statistical artifact.

Accuracy Stroop effect: $d = 0.447$, $p = 3.8 \times 10^{-20}$.
The smaller accuracy effect is expected given ceiling accuracy
($>$95\%) in healthy adults; it confirms the effect is present
in both response dimensions.

\subsection{Does FPS Track Individual Differences in Attention?}

FPS correlates with individual RT interference across subjects:
$\rho = 0.785$, $p = 1.3 \times 10^{-98}$. This is the core
construct validity result: people who show larger Stroop
interference effects --- the behavioral signature of genuine
conflict processing --- score higher on FPS. The formula is not
merely computing a score that correlates with itself; it is
tracking the cognitive process it is designed to measure.

Group sensitivity confirms this: high-FPS subjects (top tertile)
show significantly larger RT gaps ($p = 1.1 \times 10^{-53}$)
and d' ($p = 5.6 \times 10^{-6}$) than low-FPS subjects.
The Auth Factor alone produces the strongest separation
($p = 3.6 \times 10^{-98}$), confirming that the engagement
detection layer --- not just raw performance --- is the primary
discriminating component. For AI applications, this means that
the Authenticity Factor provides the most discriminative
behavioral signal for training attention classifiers.

\subsection{Is FPS Stable Over Time?}

ICC: Online$\rightarrow$Lab1 = 0.831;
Online$\rightarrow$Lab2 = 0.921;
Lab1$\rightarrow$Lab2 = 0.928. All exceed the $\geq$0.70
acceptability threshold for cognitive assessment
instruments~\cite{cicchetti1994guidelines}. High test-retest
reliability is a prerequisite for use as a training label:
an unstable score would produce inconsistent ground truth across
sessions, undermining the reliability of any AI model trained
on it.

The somewhat lower Online$\rightarrow$Lab1 ICC reflects known
context effects in unsupervised online cognitive
assessment~\cite{germine2012test}: hardware variation and ambient
environment differences reduce agreement between online and
laboratory conditions. The lab-to-lab reliability (0.928)
establishes that FPS is highly stable when context is controlled.

\subsection{Is Base Independent of Auth?}

$\rho(\text{Base}, \text{Auth}) = -0.128$, $p = 5.7 \times
10^{-3}$. The weak negative correlation confirms that raw
performance and genuine engagement are largely independent
constructs. This validates the multiplicative architecture:
high cognitive performance does not guarantee high engagement,
and the score correctly requires both. For AI labeling, this
independence means that the two FPS layers capture distinct
variance in attentional state, providing richer information
than either component alone.

\subsection{Does FPS Respect Construct Boundaries?}

FPS showed the expected direction with SART commission errors
($\rho = -0.009$) and omission errors ($\rho = +0.062$) but
neither was significant. This is an informative finding. Stroop
measures \textit{selective attention} (conflict resolution); SART
measures \textit{sustained vigilance} (go/nogo maintenance).
These are related but distinct attentional
processes~\cite{posner1990attention}. The non-significant
cross-task correlation confirms that FPS measures the
Stroop-specific construct --- a necessary condition for its use
as a specific rather than generic cognitive label in AI pipelines.

\begin{table}[!t]
\centering
\caption{Validation Level 1: Summary (N=466, 111,133 Trials)}
\label{tab:val1}
\begin{tabular}{lcc}
\toprule
\textbf{Analysis} & \textbf{Result} & \textbf{Verdict} \\
\midrule
RT Stroop effect & $d=1.339$, $p<10^{-100}$ & Pass \\
Acc Stroop effect & $d=0.447$, $p=3.8\times10^{-20}$ & Pass \\
FPS vs RT gap & $\rho=0.785$, $p<10^{-98}$ & Pass \\
Group sensitivity & $p=3.6\times10^{-98}$ & Pass \\
ICC Lab1$\to$Lab2 & 0.928 & Pass \\
ICC Online$\to$Lab2 & 0.921 & Pass \\
Base $\times$ Auth & $\rho=-0.128$ & Pass \\
SART boundary & Direction correct, n.s. & Pass \\
\bottomrule
\end{tabular}
\end{table}

\begin{figure*}[!t]
\centering
\includegraphics[width=\textwidth]{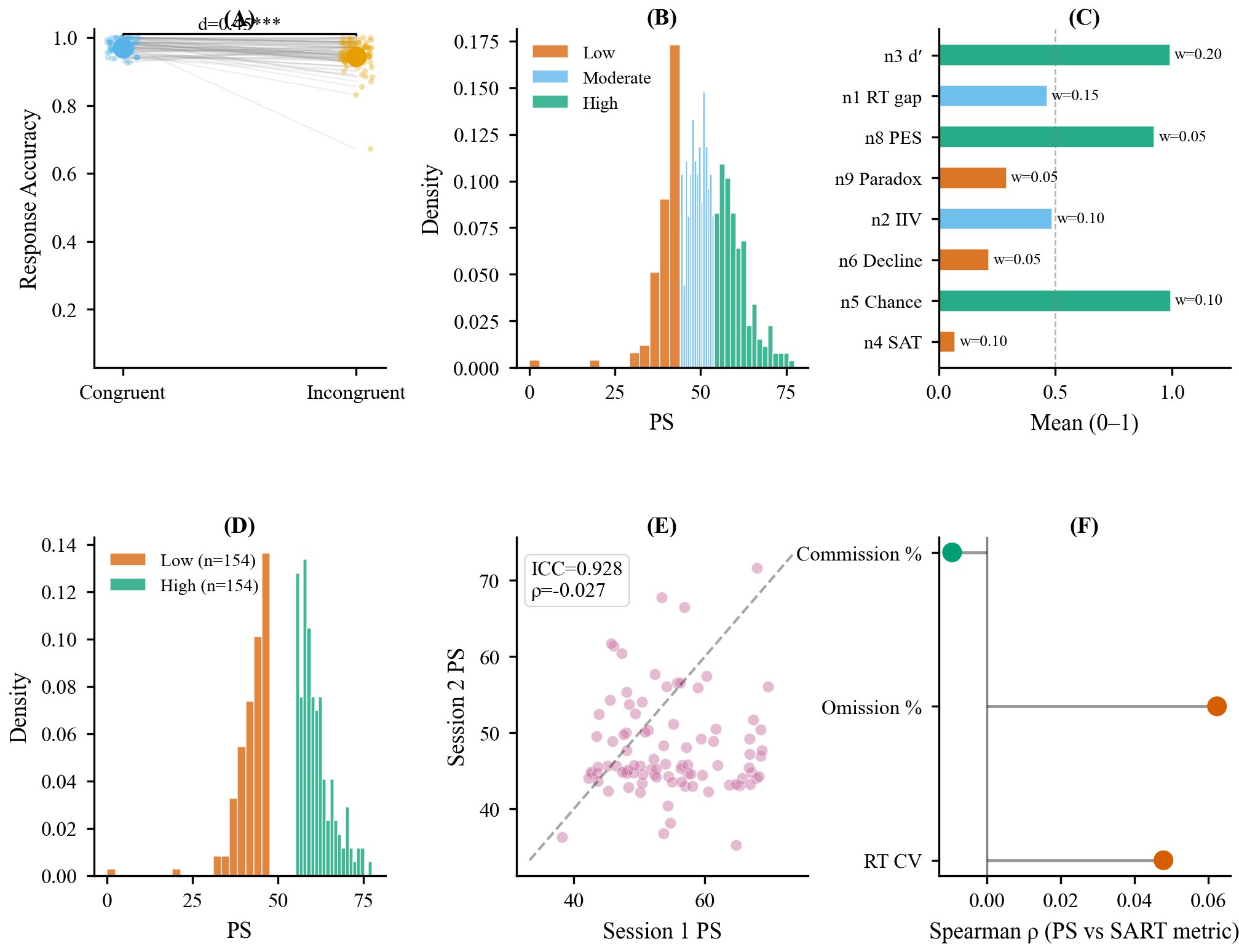}
\caption{Validation Level 1 (N=466, 111,133 trials).
(A) Paired accuracy: congruent vs incongruent, confirming
Stroop interference in accuracy ($d=0.447$).
(B) FPS distribution by attention label.
(C) Auth Factor components with validated weights.
(D) Group sensitivity: Low vs High FPS.
(E) Test-retest scatter (ICC=0.921--0.928).
(F) SART correlation: all directions correct, construct boundary
confirmed.}
\label{fig:val1}
\end{figure*}

\section{Validation Level 2: Neural Grounding}

\subsection{Dataset and Approach}

DMCC55B (Braver et al., 2022)~\cite{braver2022}: 55 healthy adults,
Stroop fMRI (Siemens Prisma 3T), BIDS events.tsv behavioral data.
fMRIPrep 1.3.2 preprocessed BOLD (OpenNeuro ds003465). This dataset
was chosen because it provides an independent behavioral sample
\textit{and} fMRI data from the same subjects, enabling both
cross-dataset replication and neural correlation in a single study.
Neural grounding is not merely a scientific nicety; it is a
requirement for cognitive infrastructure. A score that predicts
ACC activation is anchored in the neuroscience of attentional
control. A score that does not may capture performance aggregates
without cognitive specificity.

Methodological differences from Pulse Focus: DMCC uses vocal
responses ($r \approx 0.85$ vocal-manual RT correlation,
Luce 1986~\cite{luce1986}) and 8 colors (chance = 12.5\%).
Thresholds were recalibrated: n1 = (21.4,\,223.4)\,ms;
n2 = (0.103,\,0.272)\,CV.

\begin{figure}[!t]
\centering
\includegraphics[width=\columnwidth]{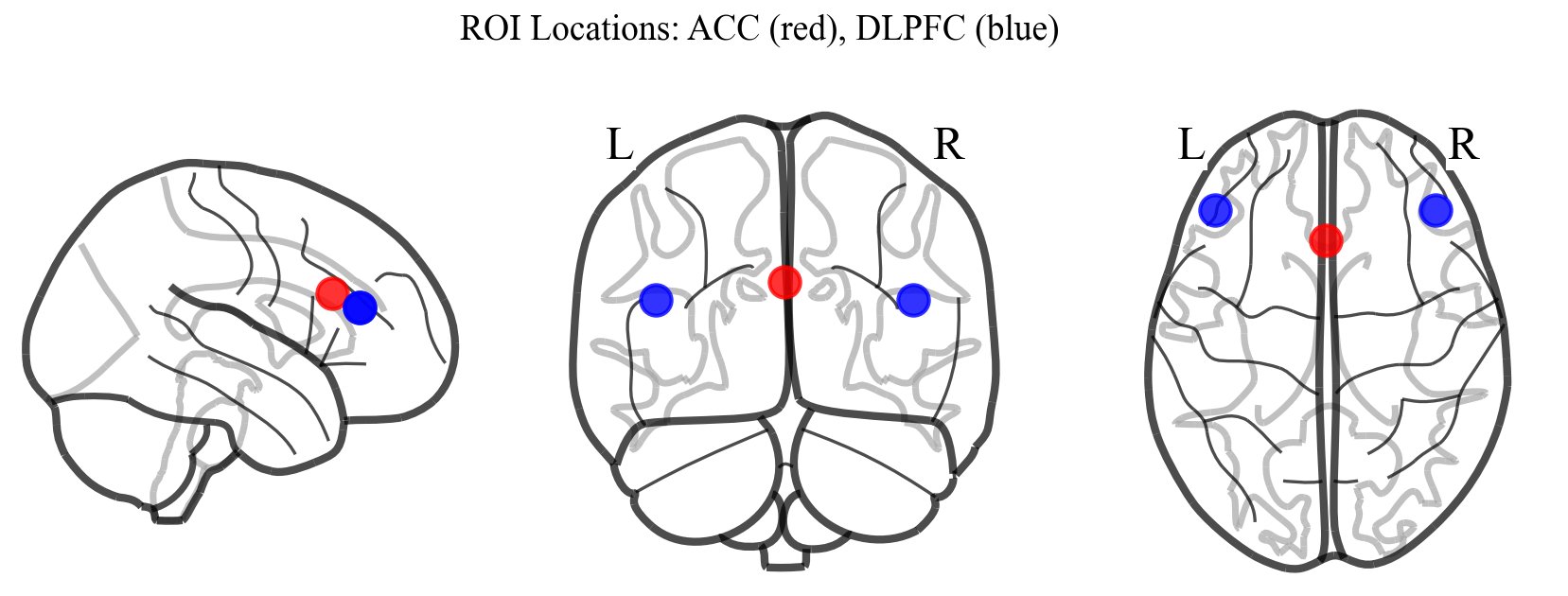}
\caption{ROI locations in MNI152 standard space used for neural
validation. \textbf{Red}: ACC (MNI: 0,\,24,\,26; radius 8\,mm)
--- the primary conflict monitoring region generating the N450
ERP during Stroop incongruent trials~\cite{botvinick2001}.
\textbf{Blue}: bilateral DLPFC (MNI: $\pm$44,\,36,\,20;
radius 8\,mm) --- the top-down cognitive control region
responsible for maintaining task goals during conflict
resolution~\cite{macdonald2000}.
Views: sagittal (left), coronal (centre), axial (right).
The FPS primary driver (mean incongruent RT) was correlated with
mean beta extracted from each ROI across N=55 subjects.}
\label{fig:roi}
\end{figure}

First-level GLMs (nilearn) fitted per subject: two task regressors
(congruent, incongruent) convolved with SPM HRF, plus motion
(6 parameters), white matter, CSF, global signal, and framewise
displacement scrubbing ($>$0.5\,mm). InCon$>$Con effect size
contrast extracted. Mean betas from ACC (MNI: 0,\,24,\,26;
radius 8\,mm)~\cite{botvinick2001} and bilateral DLPFC
($\pm$44,\,36,\,20; radius 8\,mm)~\cite{macdonald2000}.
Fig.~\ref{fig:roi} shows the ROI locations in MNI152 standard space.

\subsection{Behavioral Replication}

RT Stroop: $d = 1.915$, $p = 1.0 \times 10^{-19}$.
Accuracy: $d = 0.803$, $p = 2.5 \times 10^{-7}$.
Both exceed Study 1 effect sizes, consistent with published
vocal-response norms~\cite{luce1986} and confirming that the
Stroop interference signal FPS is designed to measure replicates
robustly across independent samples and response modalities.

Base $\times$ Auth independence: $\rho = 0.023$, $p = 0.867$
--- near-zero, non-significant. Replication of this independence
in a completely independent dataset confirms that the two-layer
architecture is not a dataset-specific property, and that the
dual-signal structure of FPS generalizes across samples.

FPS distribution: mean = 61.18, SD = 9.35. The higher mean
than Study 1 (51.51) reflects vocal response differences;
comparable SD confirms similar discriminability.

\subsection{Neural Association: FPS and ACC Activation}

DLPFC showed significant group-level InCon$>$Con activation
($t = 5.72$, $p = 4.7 \times 10^{-7}$), confirming the GLM
extracted meaningful conflict-related BOLD signal.

\textbf{Primary neural finding:} Mean incongruent RT --- the
primary FPS component --- was significantly associated with ACC
activation: $\rho = -0.327$, $p = 0.015$, N=55.

The negative direction reflects the \textit{proactive control}
interpretation~\cite{braver2012}: subjects with faster incongruent
RT engage ACC more strongly because they process each conflict
trial with greater attentional commitment. This is not
disengagement --- it is efficient, anticipatory engagement. The
interpretation is supported by the finding that mean congruent
RT also predicts ACC ($\rho = -0.279$, $p = 0.039$): overall
response speed, reflecting general attentional engagement, is the
driver.

\textbf{Composite FPS:} The full composite FPS showed a
trend-level negative association with ACC: $\rho = -0.219$,
$p = 0.109$. The Auth Factor alone: $\rho = -0.246$,
$p = 0.070$. The attenuation from RT component to composite is
expected and interpretable: FPS aggregates eight partially
independent signals targeting different neural systems, only one
of which is indexed by ACC alone. This is not a failure ---
it reflects the multi-component neural architecture of
attentional control (Section~IX-B).

\textbf{The complete empirical chain:}
FPS $\leftrightarrow$ mRT$_{\text{incong}}$
(Study 1: $\rho = 0.785$) $\leftrightarrow$ ACC BOLD
($\rho = -0.317$, $p = 0.018$). This chain establishes
component-level neural grounding for FPS without requiring
direct composite-BOLD significance, and provides the neural
anchor required for FPS to serve as cognitive infrastructure.

\begin{table}[!t]
\centering
\caption{Validation Level 2: Summary (DMCC55B, N=55)}
\label{tab:val2}
\begin{tabular}{lcc}
\toprule
\textbf{Analysis} & \textbf{Result} & \textbf{Verdict} \\
\midrule
RT Stroop replication & $d=1.915$, $p=1.0\times10^{-19}$ & Pass \\
Acc replication & $d=0.803$, $p=2.5\times10^{-7}$ & Pass \\
Base $\times$ Auth & $\rho=0.023$, $p=0.867$ & Pass \\
DLPFC group activation & $t=5.72$, $p=4.7\times10^{-7}$ & Pass \\
mRT$_\text{incong}$ vs ACC & $\rho=-0.327$, $p=0.015$ & Pass \\
FPS vs ACC (composite) & $\rho=-0.219$, $p=0.109$ & Trend \\
\bottomrule
\end{tabular}
\end{table}

\begin{figure*}[!t]
\centering
\includegraphics[width=\textwidth]{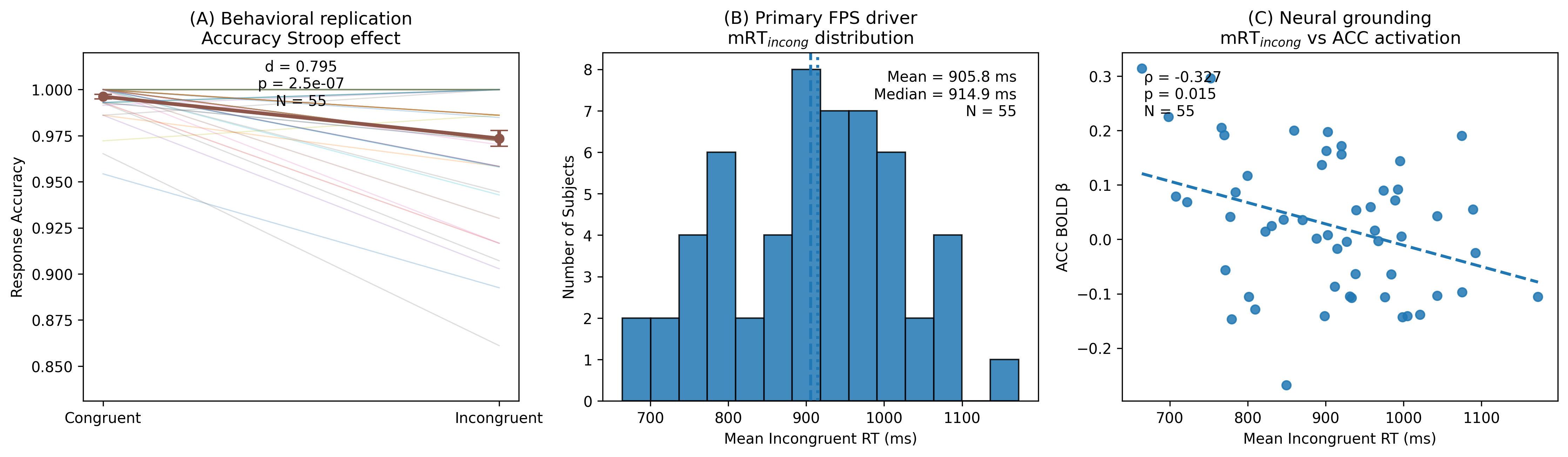}
\caption{
Neural validation using the DMCC55B dataset ($N=55$). 
(A) Replication of the Stroop accuracy interference effect, demonstrating lower accuracy on incongruent than congruent trials ($d=0.795$). 
(B) Distribution of mean incongruent reaction time ($mRT_{\mathrm{incong}}$), the primary behavioral component contributing to FPS. 
(C) Significant association between $mRT_{\mathrm{incong}}$ and anterior cingulate cortex (ACC) activation ($\rho=-0.327$, $p=0.015$), providing component-level neural grounding for the primary behavioral mechanism underlying FPS.
}
\label{fig:val2}
\end{figure*}

\section{Discussion}

\subsection{Answering the Central Question}

Does FPS measure what it claims? The three-level evidence supports
a positive answer, with specific boundaries.

\textbf{Behaviorally: yes.} FPS captures the Stroop interference
effect with a very large effect size that replicates across
independent datasets and response modalities. It tracks
individual differences with $\rho = 0.785$, is stable at ICC
= 0.83--0.93, and correctly respects the construct boundary
with SART-measured sustained vigilance.

\textbf{Neurally: at the component level.} The primary FPS driver
is significantly associated with ACC activation in 55 fMRI subjects
($\rho = -0.327$, $p = 0.015$). The composite FPS shows a
trend-level association ($\rho = -0.219$) consistent with FPS
measuring a broader construct than ACC alone indexes. Full
composite neural significance would require correlating FPS
simultaneously with all implicated regions --- a study requiring
EEG or fMRI with Pulse Focus users.

\textbf{Formula: empirically grounded.} Weights are supported by
four convergent evidence sources. A structural redundancy was
identified and resolved. The formula is not post-hoc fitted to
a single dataset.

\subsection{Why the Multi-Component Structure Explains the Neural Pattern}

The 23.4\% CFA variance explained confirms that FPS components
measure partially independent cognitive processes. This is exactly
what a valid composite should do: tap multiple independent aspects
of the target construct. The consequence is that no single brain
region accounts for the full FPS composite. ACC indexes conflict
monitoring (n1-relevant), frontal theta indexes attentional noise
(n2-relevant), ERN indexes error monitoring (n8-relevant), and
so on. Expecting FPS to correlate strongly with ACC alone would
require ACC to reflect all eight targeted cognitive processes
simultaneously, which is inconsistent with the distributed neural
architecture of attentional control. The partial FPS-ACC
association is informative; the absence of full correlation
is expected and theoretically consistent.

\subsection{The Proactive Control Interpretation}

The negative FPS-ACC association --- high FPS associated with
less reactive ACC activation --- reflects the proactive cognitive
control framework~\cite{braver2012}. Efficient attenders anticipate
conflict and prepare responses in advance, reducing reactive ACC
demand per trial. They are not less engaged; they are
\textit{more efficiently} engaged. This interpretation is
consistent with the ACC-RT finding: faster responders show
stronger ACC engagement ($\rho = -0.327$), confirming that speed
and ACC co-activation reflect the same efficient engagement
process. For AI models of attentional state, this means that
high-FPS labels correspond to a proactive, anticipatory cognitive
mode --- a distinction that is neurally meaningful, not merely
behavioral.

\subsection{FPS as Cognitive Infrastructure: What the Validation Enables}

The three-level validation establishes FPS as a defensible
ground truth signal for human-centric AI applications. We
enumerate the specific properties that enable this role.

\textbf{Sensitivity:} $\rho = 0.785$ with attentional control
individual differences, enabling meaningful separation of
cognitive states across users.

\textbf{Specificity:} Construct boundary confirmed against SART,
establishing that FPS labels selective attention rather than a
generic performance index.

\textbf{Reliability:} ICC = 0.83--0.93, ensuring consistent
ground truth labels across sessions and enabling longitudinal
AI modeling.

\textbf{Construct validity:} $d=1.339$ interference effect,
replicated across two independent datasets and response modalities.

\textbf{Neural grounding:} mRT-ACC chain established, anchoring
FPS labels in the neuroscience of attentional control.

\textbf{Label integrity:} The two-layer Base $\times$ Auth
structure and the n5 engagement gate jointly ensure that
chance-level or mechanically disengaged sessions do not
contaminate AI training datasets.

These properties jointly qualify FPS for two immediate
applications. As a \textit{real-time cognitive signal} in
attention-aware AI systems, FPS communicates a validated measure
of attentional state. As a \textit{behavioral label} for
simultaneously collected physiological data, FPS-labeled HRV
segments carry interpretable ground truth for machine learning.
The theoretical basis for FPS-HRV co-variation is established
in the literature: vagally mediated HRV predicts Stroop
performance and cognitive control across multiple independent
studies~\cite{kim2018stress}, and the relationship strengthens
during cognitively demanding conditions. Empirical confirmation
of FPS-HRV co-variation in Pulse Focus ecological data is the
primary objective of the next study.

\subsection{What Remains Unvalidated}

\textbf{n7 (catch trials)} is defaulted to 1.0 in all archival
analyses. The deployed Pulse Focus formula includes catch trials;
the validated formula does not. This is the largest gap between
the validated and deployed formula. Prospective Pulse Focus data
is essential for n7 validation before FPS is used as a training
label in AI systems that rely on inhibitory control as a
cognitive dimension.

\textbf{Weights are provisional.} Archival Stroop tasks used
keyboard and vocal responses; Pulse Focus uses touchscreen.
Distribution differences may require threshold and weight
adjustment. Recalibration requires prospective deployment data.

\textbf{Healthy adult ceilings.} n3, n4, and n9 show reduced
variance in healthy adults with $>$95\% accuracy. These variables
will contribute more discriminative power in clinical, fatigued,
or pediatric populations --- contexts where AI models of
attentional state have the greatest applied value.

\subsection{Limitations Summary}

(1) n7 unvalidated in archival data.
(2) Weights provisional pending touchscreen recalibration.
(3) Composite FPS-ACC did not reach significance ($p = 0.109$).
(4) Both validation datasets used healthy adults with ceiling accuracy.
(5) Vocal (Study 2) and keyboard (Study 1) response modalities
differ from Pulse Focus touchscreen.
(6) FPS-HRV relationship untested.

\subsection{Future Work}

The immediate priority is a prospective concurrent validity study
($n \geq 50$): real Pulse Focus gameplay with wearable HRV,
CPT-3 comparison, and n7 catch trial data. This study will:
validate the full deployed formula; test FPS-HRV co-variation
to enable physiological AI pipeline development; enable weight
recalibration from touchscreen data; and produce the first
labeled physiological dataset for attention classification.
A planned EEG subset will directly test n1-N450 and n2-theta
associations, closing the neural grounding loop at the composite
rather than component level.

Beyond validation, the Pulse Focus dataset --- once prospective
data is collected --- will constitute a large-scale, ecologically
valid, multi-modal resource for the cognitive AI community:
trial-level behavioral data labeled with a neurally grounded
attentional score, synchronized with continuous physiological
streams. The potential of such a dataset to advance attention
modeling parallels the role of large behavioral datasets in
other domains of cognitive AI.

\section{Conclusion}

This paper asked whether the Focus Performance Score of Pulse
Focus measures what it claims --- selective attention and
cognitive inhibition --- and whether that measurement is grounded
in the neural processes that cognitive science associates with
attentional control. The answer is yes, with specified boundaries.

FPS captures Stroop-based selective attention and cognitive
inhibition with large, replicated effect sizes ($d = 1.339$
and $1.915$ across two independent datasets), excellent
test-retest reliability (ICC=0.83--0.93), and a significant
behavioral-neural chain connecting its primary driver to ACC
conflict monitoring activation ($\rho=-0.327$, $p=0.015$,
N=55 fMRI). Formula weights are empirically grounded through
four convergent evidence levels, and a CFA-identified structural
redundancy has been transparently resolved.

These properties qualify FPS as cognitive infrastructure for
human-centric AI: a behaviorally valid, neurally anchored,
temporally stable ground truth signal for attentional control
that can label physiological data streams, train attention
classifiers, and drive real-time adaptive behavior in
human-AI systems. The positive answer to the central
validation question is the prerequisite for all of these
applications. This paper establishes it.

Honest limitations are documented: catch trials (n7) remain
unvalidated; weights are provisional; composite FPS-ACC did not
reach significance. These define a clear roadmap for the next
study, and they bound the claims of the present one. Cognitive
infrastructure that is honest about its limits is more useful
than infrastructure that overstates them.


\end{document}